\documentstyle[prl,multicol,aps]{revtex} 
\input epsf

\begin{document}

\draft 


\title{Supercooled Water: Dynamics, Structure and Thermodynamics}

\author{Francis W. Starr$^{1,4}$, \underline{Srikanth Sastry}$^{2~*}$, 
Francesco Sciortino$^{3}$ and H. E. Stanley$^{4}$}

\address{$^1$ Center for Theoretical and
Computational Materials Science and Polymers Division, National Institute of Standards and
Technology, \\ Gaithersburg, Maryland 20899, USA }

\address{$^2$ Jawaharlal Nehru Centre for Advanced Scientific Research, 
Jakkur Campus, Bangalore 560064, INDIA}

\address{$^3$ Dipartimento di Fisica and Istituto Nazionale
 per la Fisica della Materia, Universit\'a di Roma {\it La Sapienza},\\
 P.le Aldo Moro 2, I-00185, Roma, ITALY}

\address{$^4$ Center for Polymer Studies and Department of Physics, Boston University, Boston MA 02215, USA}

\maketitle

\begin{abstract}
The anomalous properties of water in the supercooled state are
numerous and well-known. Particularly striking are the strong changes
in dynamic properties that appear to display divergences at
temperatures close to -- but beyond -- the lowest temperatures attainable
either experimentally or in computer simulations. Recent work on slow
or glassy dynamics in water suggests analogies with simple liquids not
previously appreciated, and at the same time highlights some aspects
that remain peculiar to water. A comparison of the behavior of water
with normal liquids, with respect to its dynamic, thermodynamic and
structural changes in the supercooled regime is made by analyzing, via
computer simulations, the properties of local potential energy minima
sampled by water in supercooled temperatures and pressures.
\end{abstract} 



\begin{multicols}{2}

\section{Introduction} 

Water is among the most abundant substances on earth, and also among
the most familiar, as it forms a significant part of the natural
environment, and of living organisms. In addition to these obvious
reasons, the continued study of water by physical scientists stems
from the many remarkable and peculiar properties it possesses, some of
which are part of the popular lore (such as the fact that liquid water
is heavier than ice near ambient conditions). In the crystalline
state, water exists in as many as twelve, possibly fourteen, forms of
ice.  Echoes of this {\it polymorphism} are to be found in the {\it
amorphous} state as well.  When prepared as an {\it amorphous} solid,
or glass, by ultrafast cooling or by depositing vapor on a cold
substrate, water is found in two forms, a high density and a low
density form \cite{mishima}. 

A substantial fraction of the study of liquid water has focussed on
its properties in the supercooled state\cite{angell83}, where its
unusual properties are amplified. Analyzing a range of dynamic and
thermodynamic quantities, Speedy and Angell\cite{speedy-angell} showed
that these quantities appeared to diverge at a finite temperature in a
power law fashion, leading to the hypothesis of a thermodynamic
singularity at $T_s = 228 K$. Computer simulation studies and analysis
of simplified models have in recent years led to the proposal of
two other possibilities, namely the existence of a liquid-liquid
critical point at low temperatures\cite{poole}, and the scenario wherein no
thermodynamic singularities need be invoked to explain the anomalies
observed in water\cite{sastry96}.

However, there has been a renewed focus recently on the idea that
regardless of a thermodynamic singularity, strong changes in both the
thermodynamic and dynamic properties of supercooled water may be
expected in the vicinity of
$T_s$\cite{sciortino,itoet,starr}. Evidence from computer simulations
suggests that the strong temperature dependence of dynamical
quantities, and the power law temperature dependence, may be explained
as manifestations of an avoided dynamical singularity described by the
mode coupling theory of the glass transition\cite{sciortino}. Analysis
of experimental data\cite{itoet,starr} appears to indicate that 
the rapid changes in the entropy of supercooled water must cross over
to slower changes in the vicinity of $T_s$, and correspondingly, the heat
capacity must display a maximum. It has been argued\cite{itoet,starr}
that upon crossing this temperature, water changes character from being a
very ``fragile'' liquid at higher temperatures, to a ``strong'' liquid
at lower temperatures\cite{fragility}. Specifically, the stronger than
Arrhenius temperature dependence of relaxation times seen at
temperatures higher than $T_s$ is argued to cross over to Arrhenius
dependence at lower temperatures (the experimental situation is
ambiguous at present, with measurements in Ref.\cite{bruce} arguing
in favor of continuous change down to the glass transition temperature
$\sim 135 K$).

Among the crucial issues one must address in understanding the above
possibilities is the nature of structural change that takes place as
liquid water is supercooled. In particular, the nature of structural
change that accompanies a fragile to strong crossover must be
understood.  In this paper we present a preliminary report of
investigations to address these questions, which focus on analyzing
the properties of local potential energy minima or ``inherent
structures'' sampled by a simulation model of water. Similar
investigations have recently proved a fruitful approach to studying
changes in liquid state properties under supercooling of model atomic
liquids as well as liquid
water\cite{sastry98,fs,scala,ruocco,heuer,roberts}. In particular, we
focus on the comparison of changes in dynamics with changes in the
energies of the inherent structures sampled, and structural change as
evidenced by changes in the nearest neighbor geometrical properties.
A specific structural feature we consider is the fraction of water
molecules that are four-coordinated and those with five or greater
other molecules in their first neighbor shell. These latter
``bifurcated bond'' arrangements have previously been shown to
facilitate structural rearrangement and hence faster
dynamics\cite{sciortino2}.
\begin{figure}
\hbox to\hsize{\epsfxsize=1.0\hsize\hfil\epsfbox{
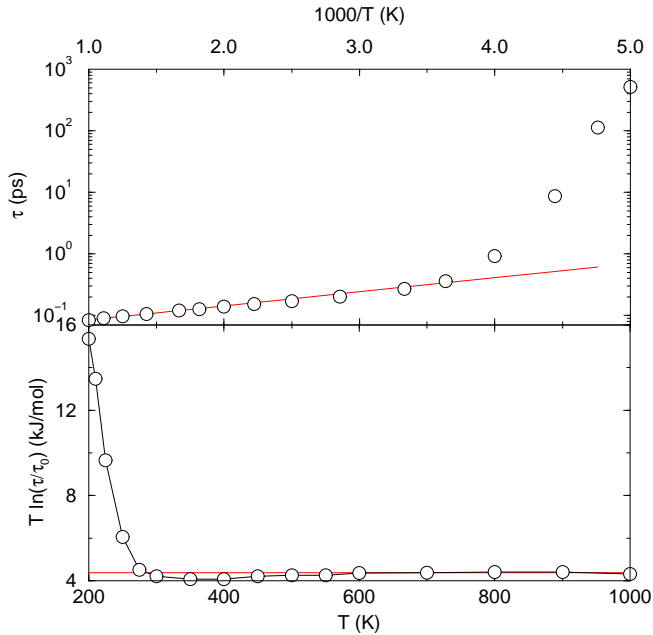}\hfil}
\caption{\narrowtext Relaxation times: (a) Shown on a logarithmic scale {\it vs.} inverse 
temperature (Arrhenius plot). Deviations from Arrhenius behavior is
observed at low temperatures. (b) Transformed to yield a constant value if 
Arrhenius behavior is obeyed. }
\label{fig1}
\end{figure}

\section{Simulation Details} 

We perform molecular dynamics (MD) simulations of 216 water molecules
interacting via the SPC/E pair potential~\cite{spce}. The simulations 
we describe here are for a fixed density of $\rho = 1.0 g/cm^3$, for 
temperatures ranging from $210 K$ to $700 K$. For $T \le
300$~K, we simulate two independent systems to improve statistics, as
the long relaxation times make time averaging more difficult. Full 
details of the simulation protocol used are described in \cite{simdet}. 
At the studied density, the mode coupling singular temperature, $T_c$ 
has been estimated to be $T_c = 194 K  \pm 1$\cite{simdet}. 

Inherent structures are obtained, with a sample of 100
equilibrated liquid configurations as starting configurations, by
performing a local minimization of the potential energy, using
conjugate gradient minimization.  The iterative procedure for
minimizing the energy is performed until changes in potential energy
per iteration are no more than $10^{-15} kJ/mol$.  A normal mode analysis
confirms that the potential energy changes with positive curvature
along all directions away from the minimum energy configurations so obtained. 

We calculate the oxygen-oxygen pair correlation function for both the
equilibrium liquid configurations, and the inherent structures. The
integrated intensity under the first peak of the pair correlation
function (upto the cutoff radius $0.31 nm$) yields the coordination
number. Dynamics is probed by calculating the self intermediate
scattering function at wavevector 18.55 nm$^{-1}$, {\it i.e.} the
Fourier transform of the self van Hove self correlation function
\begin{equation} 
G_s(r,t) = {1 \over N} \sum_{i = 1}^{N} \left<\delta(|{\bf r}_i(t)-{\bf r}(0)| - r)\right>. 
\end{equation}
Stretched exponential fits of the self intermediate scattering
function yield relaxation times $\tau$ as a function of
temperature. Average values of the inherent structure potential energy
and pressure are also calculated.
\begin{figure}
\hbox to\hsize{\epsfxsize=1.0\hsize\hfil\epsfbox{
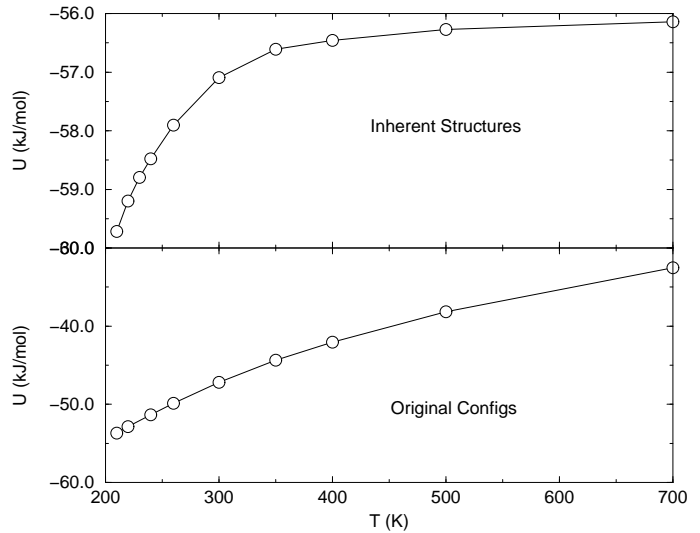}\hfil}
\caption{\narrowtext (a) Average inherent structure energies. (b) Equilibrium averaged potential energies.}
\label{fig2}
\end{figure}

\section{Results} 

Figure 1(a) shows the relaxation times  plotted on a logarithmic scale 
against inverse temperature. Clear deviations from the high temperature Arrhenius behavior, 
\begin{equation}
\tau = \tau_0 exp(-E/k_B T)
\end{equation} 
(where $E$ is a constant) is observed for $T \lesssim 275 K$.  This is
more clearly seen by plotting $T~ ln(\tau/\tau_0)$ (where $\tau_0$ is
obtained from an Arrhenius fit of the highest $5$ temperatures), as
shown in Fig. 1(b), which shows strong deviation of $T~
ln(\tau/\tau_0)$ below $T \sim 275 K$ indicating that the temperature
dependence of relaxation times at low temperatures can no longer be
described by the Arrhenius form.

Figure 2 (a) shows the average inherent structure energies as a
function of temperature. For comparison, the equilibrium potential
energy is shown in Figure 2(b). Inherent structure energies change
continuously, with modest temperature dependence at high temperatures
to substantial temperature dependence at lower temperatures. In
particular, in the range of temperatures where the relaxation times
begin to display non-Arrhenius behavior, the inherent structure
energies show considerable temperature dependence. This behavior is 
analogous to that found for a model atomic fluid\cite{sastry98}.
\begin{figure}
\hbox to\hsize{\epsfxsize=1.0\hsize\hfil\epsfbox{
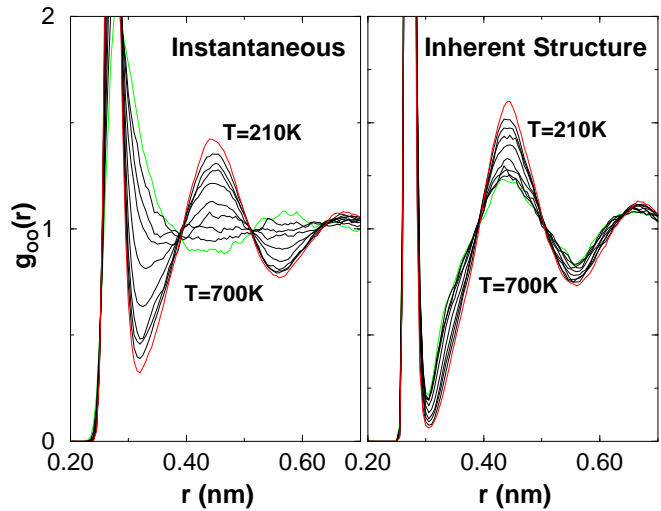}\hfil}
\caption{\narrowtext Oxygen-oxygen pair correlation function, for (a) equilibrated liquid configurations, and (b) inherent structures. Upon lowering temperature, the first peak in both cases becomes sharper, and the intensity between the first and second peaks decreases. The smaller changes seen in the case of 
the inherent structures offers an estimate of that part of the change due to
configurational change upon cooling, as opposed to thermal effects.}
\label{fig3}
\end{figure}

Figure 3 shows the oxygen-oxygen pair correlation function, (a) for
the equilibrated liquid and (b) for the inherent structures. In
contrast to the model atomic liquid studied in \cite{sastry98,fs},
both pair correlation functions show marked temperature dependence, at
fixed density. In both Fig.s 3(a) and 3(b), the first peak of the pair
correlation function becomes sharper upon decreasing the
temperature. Further, there is a systematic reduction of the intensity
between the first and second neighbor peaks.  The integrated intensity
under the first peak, which gives the average number of neighbor
molecules in the first coordinate shell, approaches the value of $4$
from higher values, as the temperature decreases.  This implies that
the configurations sampled by the liquid come closer to that of a
four-coordinated random network.
\begin{figure}
\hbox to\hsize{\epsfxsize=1.0\hsize\hfil\epsfbox{
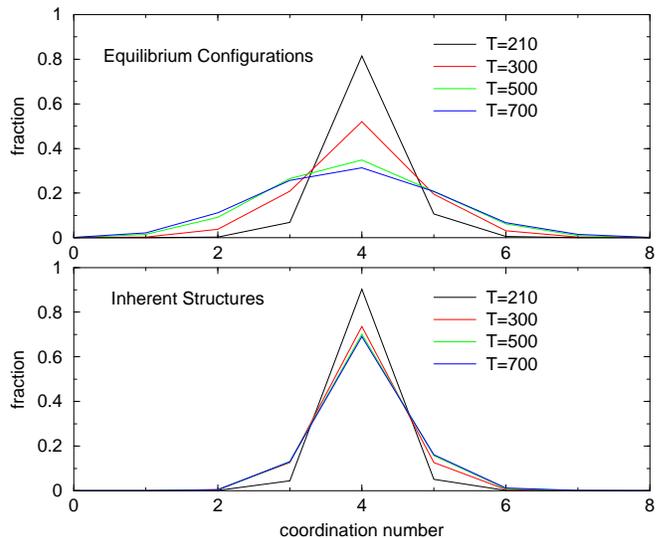}\hfil}
\caption{\narrowtext Distribution of the coordination number of molecules: (a) Equilibrated configurations, and (b) Inherent structures. In both cases, the
distribution becomes more narrowly distributed around four-coordination. }
\label{fig4}
\end{figure}

To quantify this further, Fig. 4 shows the histogram of the
fraction of molecules with a given coordination number. For
equilibrated configurations, the histogram changes from a rather broad
one to one that is peaked around the value $4$ as the temperature
decreases.  The same trend is visible for the inherent structures,
although even at high temperatures, the distribution is quite narrowly
peaked around the value $4$. Such a comparison permits us to make a
separation between deviations from four-coordination arising from 
thermal agitation\cite{fhs}, and that arising from configurational change. 
\begin{figure}
\hbox to\hsize{\epsfxsize=1.0\hsize\hfil\epsfbox{
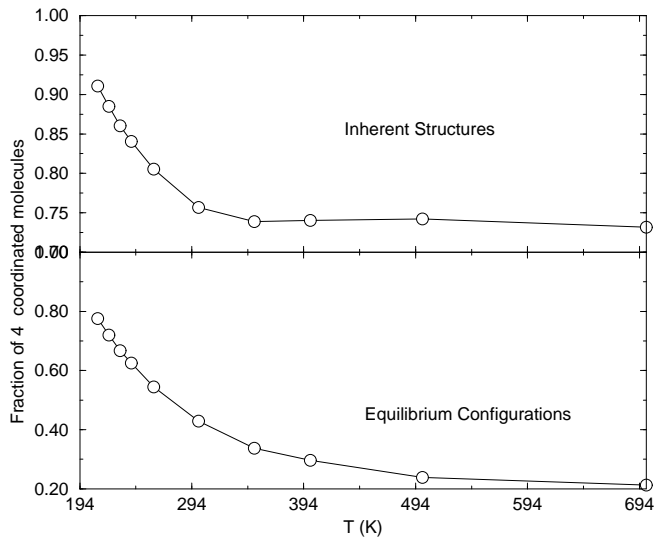}\hfil}
\caption{\narrowtext Fraction of four coordinated water molecules: (a) Equlibrated configurations, and (b) Inherent structures}
\label{fig5}
\end{figure}

From data in Fig. 3 and 4, we calculate the fraction of molecules that
are four-coordinated, and those that have five or higher coordination,
{\it i. e.} which have bifurcated bonds.  Fig. 5 shows the temperature
dependence of the fraction of four-coordinated water molecules. For
both the equilibrated configurations and inherent structures, this
fraction approaches $one$ as $T \rightarrow T_c$. As with deviations
from Arrhenius behavior, the range of temperatures displaying
substantial increase in this fraction also shows substantial
temperature dependence of the average inherent structure energies.

Figure 6 shows the fraction of molecules with five or higher coordination. 
Complementary to the variation of the fraction of four coordinated molecules, 
this fraction approaches values very close to zero as $T_c$ is approached. 
\begin{figure}
\hbox to\hsize{\epsfxsize=1.0\hsize\hfil\epsfbox{
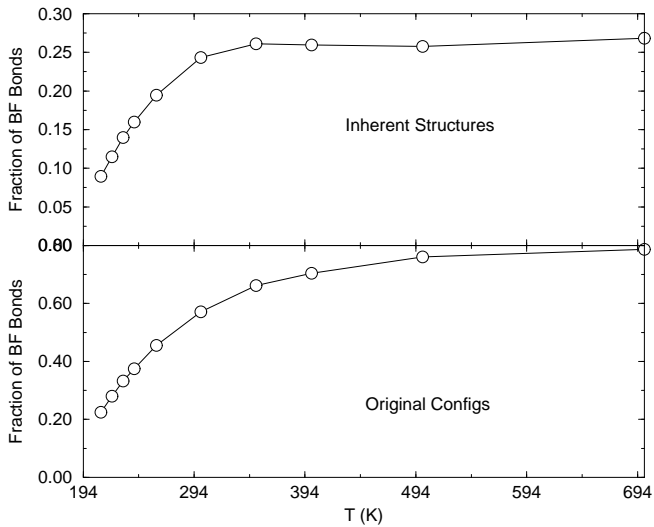}\hfil}
\caption{\narrowtext Fraction of five or higher coordinated water molecules: (a) Equlibrated configurations, and (b) Inherent structures}
\label{fig6}
\end{figure}

\section{Summary}

	We have presented simulation results that demonstrate the significant
correlations between relaxation times, average energies of local
potential energy minima sampled, and structural features for a
simulation model of water. In particular, as the mode coupling
temperature $T_c$ estimated for this model liquid at the studied
density is approached, the structure of the liquid appears to approach
that of a four-coordinated network, free of defects to the extent
permissible by the bulk density of the liquid. Such a structural
change can potentially explain the speculated dramatic changes both in
the dynamics and thermodynamic properties of water across this
crossover temperature.  Below the crossover temperature, no further
structural arrangement may be expected, and the configurational
entropy of the liquid may become `frozen in' at the value that corresponds
to the random tetrahedral network (plus the residual defects that
may be present at concentrations varying with density). Thus the
rate of change in entropy of the liquid would change substantially
near the crossover temperature. Similarly, because of the significant
temperature dependence of the fraction of bifurcated bonds -- which
facilitate structural rearrangement -- above the crossover
temperature, and the relative constancy below, the temperature
dependence of the dynamical properties may also be expected to show a
corresponding crossover. Further work is in progress to strenghten these
notions. 

We acknowledge useful discussions with Prof. C. A. Angell. FS
acknowledges INFM-PRA and MURST-PRIN, and HES acknowledges NSF, for
financial support.

\end{multicols}

\end{document}